# Histo-blood group glycans in the context of personalized medicine

Viktoria Dotz[a,b*], Manfred Wuhrer[a,b]

[a] Division of BioAnalytical Chemistry, VU University Amsterdam, Amsterdam, The Netherlands;

[b] Center for Proteomics and Metabolomics, Leiden University Medical Center, Leiden, The Netherlands;

* Corresponding author: Viktoria Dotz; Tel.: +31-71-5268701; e-mail: v.dotz@lumc.nl or v-dotz@t-online.de; postal address: Leiden University Medical Center, Postbus 9600, 2300RC Leiden, The Netherlands

Abbreviations:

BG, blood group; CA, cancer antigen; FORS, Forssman; FUT, fucosyltransferase (gene); GLOB, Globoside; GSL, glycosphingolipid; ISBT, International Society for Blood Transfusion; Le, Lewis; MS, mass spectrometry; RBC, red blood cell; Se, Secretor; sICAM-1, soluble Intercellular Adhesion Molecule 1

## Abstract

Background: A subset of histo-blood group antigens including ABO and Lewis are oligosaccharide structures which may be conjugated to lipids or proteins. They are known to be important recognition motifs not only in the context of blood transfusions, but also in infection and cancer development.

Scope of review: Current knowledge on the molecular background and the implication of histo-blood group glycans in the prevention and therapy of infectious and non-communicable diseases, such as cancer and cardiovascular disease, is presented.

Major conclusions: Glycan-based histo-blood groups are associated with intestinal microbiota composition, the risk of various diseases as well as therapeutic success of, e.g., vaccination. Their potential as prebiotic or anti-microbial agents, as disease biomarkers and vaccine targets should be further investigated in future studies. For this, recent and future technological advancements will be of particular importance, especially with regard to the unambiguous structural characterization of the glycan portion in combination with information on the protein and lipid carriers of histo-blood group-active glycans in large cohorts.

General significance: Histo-blood group glycans have a unique linking position in the complex network of genes, oncodevelopmental biological processes, and disease mechanisms. Thus, they are highly promising targets for novel approaches in the field of personalized medicine.

# 1 Introduction

More than a century after the discovery of the ABO blood group (BG) by Karl Landsteiner [1], currently, over 30 different BG systems and more than 300 recognized BG antigens are defined by the International Society for Blood Transfusion (ISBT) [2]. A recent overview is provided on the society's website [3]. The different BG antigens evolve from genetic polymorphisms of red blood cell (RBC) surface molecules, most of which are peptides, and some are carbohydrates, such as ABO antigens [3]. However, BG antigens including ABO are not only expressed on RBCs, but are also present in many tissues [4]. Therefore, these histo-BG antigens appear to be relevant not only in transfusion medicine, but also in transplantation [5]. Moreover, histo-BG antigens may also occur in glandular secretions. For example, ABO and Lewis antigens are found on saliva mucins, and free oligosaccharides are found in milk and urine [6–9]. Structurally, these particular histo-BGs are glycans, defined as "*any sugar or assembly of sugars, in free form or attached to another molecule*" [10].

Glycans in general are known to be "*directly involved in the pathophysiology of every major disease*", and it has been concluded that "*knowledge from glycoscience will be needed to realize the goals of personalized medicine and to take advantage of the substantial investments in human genome and proteome research and its impact on human health*" as stated by a recent report from the US National Academies [11].

Although the lack of expression of BG antigens is not directly resulting in disease, as known so far, the presence or absence of certain BG antigens on RBC surfaces or in other tissues or bodily fluids has been found to be associated with susceptibility to various diseases, beyond their recognized role in incompatibility reactions during transfusion, transplantation or pregnancy. Largest body of evidence for the carbohydrate-based histo-BGs ABO, Lewis (Le) and Secretor (Se) is available particularly in the context of infectious diseases and cancer (reviewed in [12–16]). Nevertheless, histo-BG glycans are far from being exploited for diagnostic or therapeutic applications, apart from the prominent exception of the cancer antigen (CA) 19-9, i.e. sialyl-$Le^a$ [17].

Glycan-based BGs, i.e. ABO, P1PK, Le, H and Se, Ii, Globoside (GLOB), Forssman (FORS), and the high-incidence antigen $Sd^a$, altogether represent more than 20 distinct antigenic structures. An overview of the glycan-based BG systems and the associated RBC phenotypes with their distributions among populations is given in Table 1. In this review, the potential use of these glycan-based histo-BGs in the context of personalized medicine will be discussed as well as the technology suitable for determining them in a research as well as clinical setting.

**Table 1 Glycan histo-blood groups (BG) and responsible genes with blood type distributions among populations**

| BG [1] | Gene [2] | Glycosyltransferase [2] | Carrier | Antigens [3] | RBC Type | Frequencies of blood types in % [4] | | | |
|---|---|---|---|---|---|---|---|---|---|
| | | | | | | Caucasians | China | US blacks | Reference |
| ABO | ABO | (Inactive) | | (H) | O | 39 | (34) | 49 | [12,18,19] |
| | | Histo-blood group ABO system transferase (**A transferase**; alpha 1-3-N-acetylgalactosaminyl-transferase; A3GALNT) | GSL, N-/O-glycoprotein, free glycan | A; A1; A2 | A | 42 | (29) | 27 | |
| | | Histo-blood group ABO system transferase (**B transferase**; alpha 1-3-galactosyltransferase; A3GALT1) | | B | B | 13 | (28) | 20 | |
| | | **A transferase** and **B transferase** | | A; A1; A2; B | AB | 6 | (9) | 4 | |
| LE | FUT3 | (Inactive) | | | Le(a-b-) | 6 | 9 | 22 | [12,20] |
| | | Galactoside 3(4)-L-fucosyltransferase (Lewis FT; fucosyltransferase 3; CD174); **Le-FUT** | GSL, N-/O-glycoprotein, free glycan | Lea | Le(a+b-) | 22 | 0 | 23 | |
| | | | | Leb; (Lea) | Le(a-b+) [5] | 72 | 71 | 55 | |
| | | | | Lea; (Leb) | Le(a+b+) | 0 | 20 | 0 | |
| Se [6] | FUT2 | (Inactive) | | | (not applicable) [6] | (20) | | | [21,22] |
| | | Galactoside 2-alpha-L-fucosyltransferase 2 (Alpha(1,2)FT 2; Secretor factor); **Se-FUT** | GSL, N-/O-glycoprotein, free glycan | Se (Type 1 H) | | (80) | | | |
| H | FUT1 | (Inactive) | | | Bombay or para-Bombay | | rare | | [12] |
| | | Galactoside 2-alpha-L-fucosyltransferase 1 (fucosyltransferase 1); **H-FUT** | GSL, N-/O-glycoprotein | H | H | | almost 100% | | [12] |
| I | GCNT2 | (Inactive) | | i | i | | rare (adults) | | [23] |
| | | N-acetyllactosaminide beta-1,6-N-acetylglucosaminyl-transferase, isoform A (I-branching enzyme) | | I; i | I | | almost 100% (adults) | | |

[1] According to [3]
[2] According to the HUGO Gene Nomenclature Committee at the European Bioinformatics Institute (genenames.org) in case of gene names and/or the recommendations in UniProtKB with a selection of alternative names in parentheses in case of glycosyltransferase names. Short names as used in this article are in bold.
[3] Antigens on red blood cells (RBC), tissues, or in secretions markedly elevated or specific to a certain blood type as compared to the other phenotypes within a BG. List is not extensive, i.e. not including various combinations/extensions of the respective antigens.
[4] Both genotypic and red blood cell phenotypic determinations were included here and genotyping data are shown in parentheses.
[5] Le(a-b+) phenotype only in combination with active FUT2 gene.
[6] Se-gene encoded FUT2 is not expressed in RBCs, and therefore does not represent a classic blood group, but provides Type 1 H antigens as precursors for BG antigens in secretions and tissues.

| | | | | | | | | | |
|---|---|---|---|---|---|---|---|---|---|
| **P1PK** | A4GALT | (Inactive) | | PX2 | p | | rare | | [12] |
| | | Lactosylceramide 4-alpha-galactosyltransferase (Gb3 synthase; P1/Pk synthase) | GSL | P1; Pk; (NOR) [7] | P1 | 79 | 27 | 94 | |
| | | | | Pk; (NOR) [7] | P2 | 21 | 73 | 6 | |
| **GLOB** | B3GALNT1 | (Inactive) | | | Pk | | rare | | [12,24] |
| | | UDP-GalNAc:beta-1,3-N-acetylgalactosaminyltransferase 1 (globoside synthase) | GSL | P | | | high incidence | | |
| **FORS** | GBGT1 | (Inactive) | | | | | almost 100% | | [12,24] |
| | | Globoside-3-alpha-N-acetyl-D-galactosaminyltransferase (Fs synthase) [8] | GSL | FORS1 | $A_{pae}$ | | rare | | |
| **Sid** | B4GALNT2 [9] | (Inactive) | | | $Sd^{a-}$ | | <10% | | [25] |
| | | Beta-1,4 N-acetylgalactosaminyltransferase 2 | N-glycoprotein | $Sd^{a}$ | $Sd^{a+}$ | | high incidence | | [26] |
| **(no. 209)** | ST3GAL2 [9] | CMP-N-acetylneuraminate-beta-galactosamide-alpha-2,3-sialyltransferase 2 | GSL | LKE | LKE | | almost 100% | | [12,24] |

[7] NOR antigens expressed if a rare variant of A4GALT gene is present
[8] Due to its novelty no UniProt entry exists for a human GBGT1-encoded Fs synthase. Nomenclature was used according to [27].
[9] Genetic background not yet completely understood.

## 2 Structural basis of glycan histo-blood groups

Glycan histo-BG antigens can occur in various types of glycoconjugates, i.e. on glycosphingolipids (GSLs), N- and O-glycans of cell membrane-bound or secreted glycoproteins and mucins, or on free oligosaccharides from milk or urine (Table 1, Fig. 1, [23,28]). As an example, ABO activity on erythrocytes was found to originate from glycoproteins (65–75%), polyglycosylceramides (10–15%) and from other glycoconjugates (10%) [29].

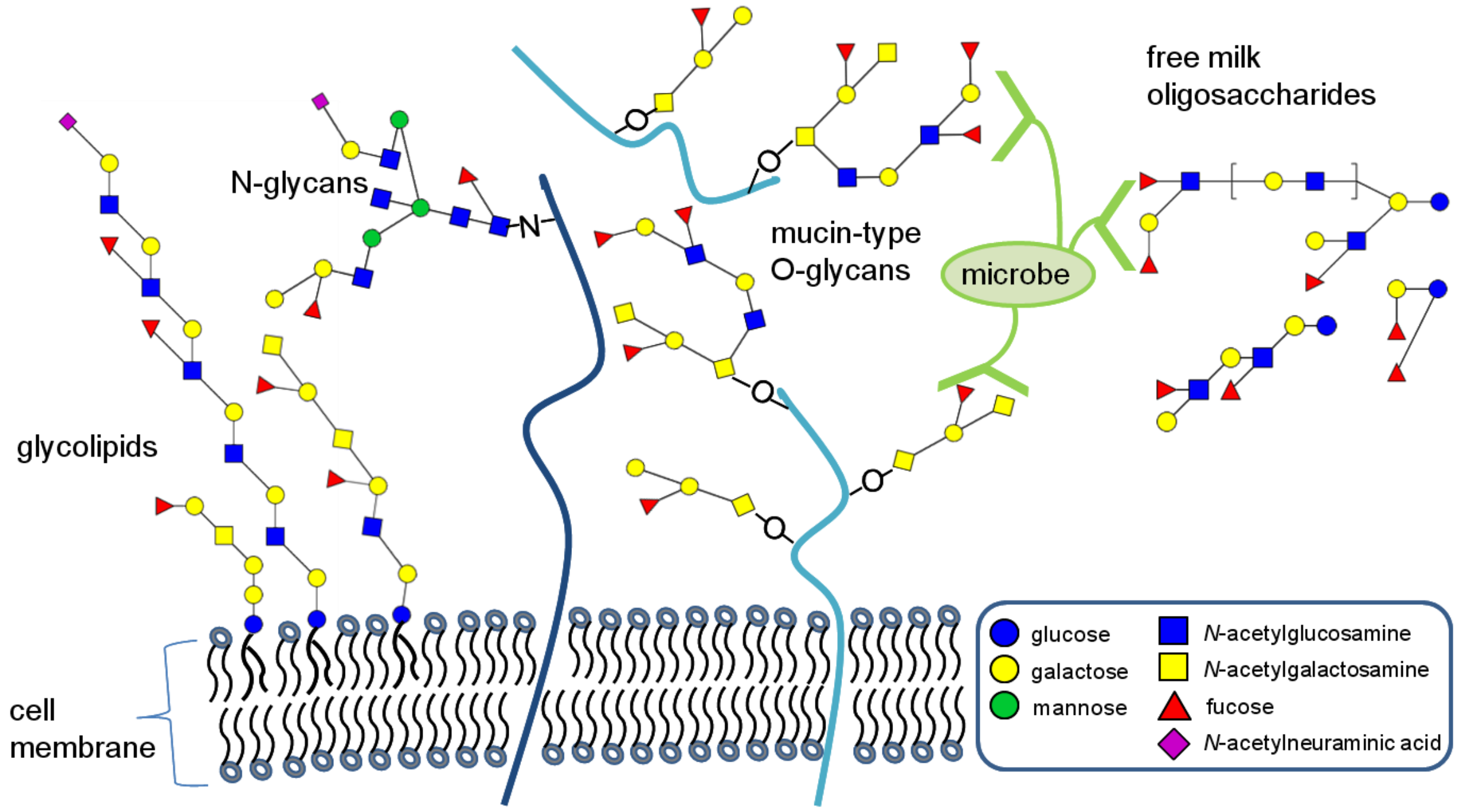

**Fig. 1 Examples of human histo-blood group glycans and their carriers on cell surfaces and in body fluids. Histo-blood group glycans decorate various core glycans attached to lipids (black waves) and proteins (blue lines) anchored in the cell membrane, or to secreted glycoproteins/mucins or free oligosaccharides as found in large quantities in human milk. Blood group antigens are found on both N- and O-glycans of proteins. Microbial receptors can recognize various histo-blood group antigens and can thereby attach to the host's epithelial surfaces. Alternatively, soluble glycans can serve as decoy receptors for pathogens.**

Even when regarding the glycan portion only, a diversity of precursor oligosaccharides together with the various antigenic determinants potentiates the overall structural diversity of histo-BG-active compounds. For instance, ABO determinants are found on type 1 and 2 chains of N-/O-glycans attached to proteins or (*neo*)lacto-series GSLs (Fig. 2A and 2B). Furthermore, they decorate O-linked type 3 chains on mucins that are structurally identical to the so-called T antigen (Fig. 2C). Type 4 chains bearing ABO epitopes are part of globo- and ganglio-series GSLs (Fig. 2C and Fig. 3; comprehensive reviews in [28,30–32]). The expression of histo-BG antigens and their precursors is tissue-specific and has been associated with the embryologic origin of a tissue and the degree of differentiation of the respective cells within a tissue [4,33–35]. In the following, the minimal structural features of glycan histo-BG epitopes are briefly described, and a summary of the relevant genes, enzyme names and products is given in Table 2. For a more extensive overview on the genetic, biochemical, epidemiological, and historical aspects of those the reader is referred to literature [4,12,23,24,26,28,36–40].

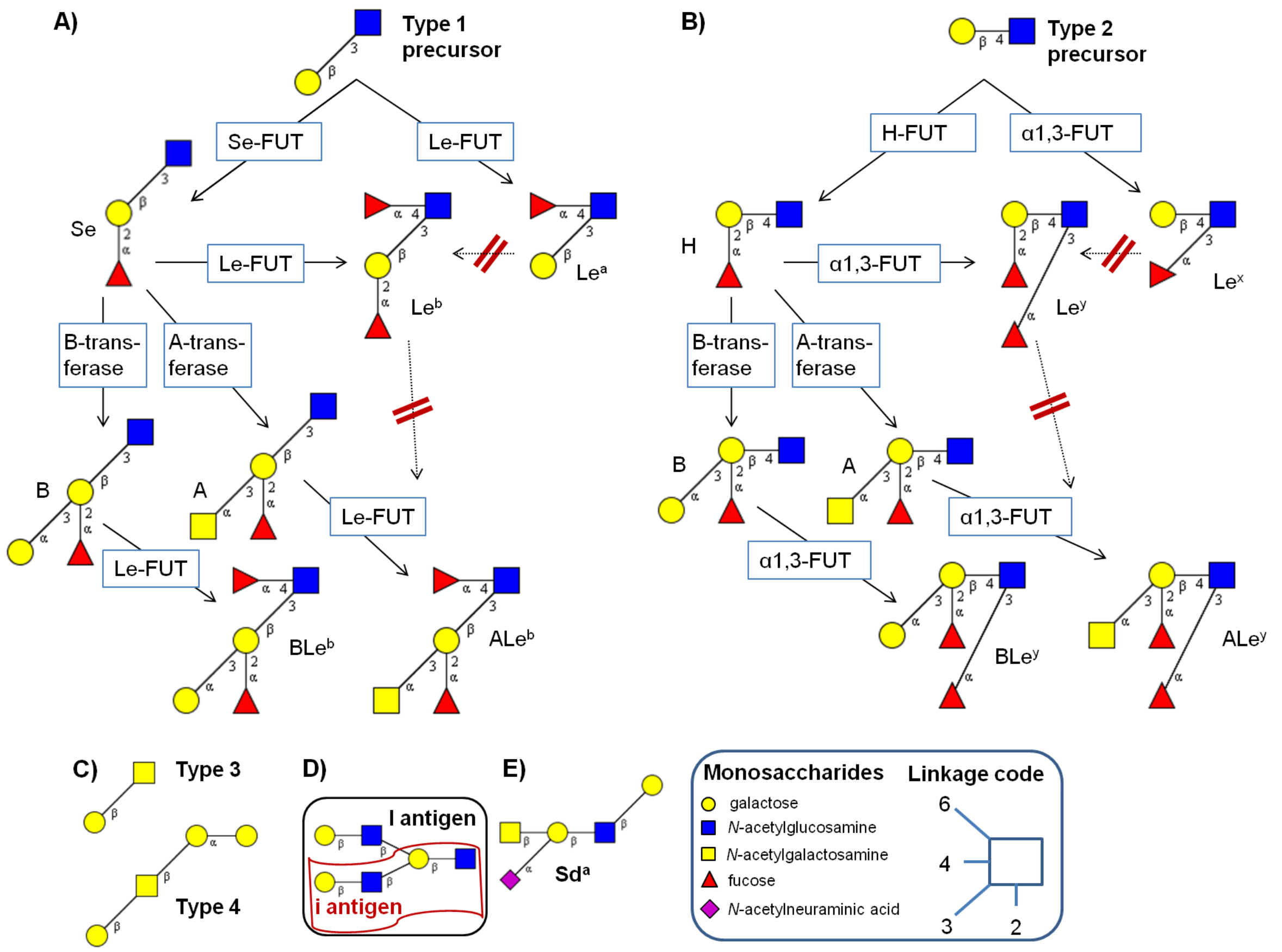

**Fig. 2 Antigenic structural motifs of the histo-blood groups ABO and Lewis (Le) with their precursors. The interaction of glycosyltransferases acting on type 1 and type 2 precursors results in ABO and Lewis a/b (A) or x/y structures (B), respectively. Mucin-type 3 (T antigen) and glycosphingolipid-based type 4 structures are also precursors to ABO and Lewis structures (C). The Ii blood group is determined by linear (i) or beta6-branched (I) polylactosamine type 2 chains (D). Sda antigenic determinant (E). FUT, fucosyltransferase; Se, secretor. Dashed, red-crossed arrows indicate inadmissible reactions.**

## 2.1 Ii histo-blood group

Ii epitopes are based on repeating units of either linear or β1,6-branched N-acetyllactosamine chains (Fig. 2D) [41–43]. The name of the blood group emerged from an abbreviation of ‘individuality’ and thus represents an upper and a lower-case letter ‘i’. Both structures are ubiquitously expressed, with the exception of the very rare adult i phenotype lacking branched I structures either on erythrocytes only or even tissue-wide, depending on the type of mutation [23,44].

Similar to other polylactosamines, Ii structures are substrates to various glycosyltransferases, including sialyltransferases and ABO and Le BG transferases as described in the following (for review see [32,45]).

## 2.2 ABO, H and Lewis histo-blood groups

The biosynthesis of the antigens from ABO and Le histo-BGs is closely related, although the responsible glycosyltransferases are expressed from several independent genes: ABO gene on chromosome 9, H-gene (FUT1), Se-gene (FUT2), and Le-gene (FUT3) on chromosome 19 [34,37]. For abbreviated as well as full names including linkage specificities and responsible genes of the fucosyltransferases (FUT), see Table 1. Starting from type 1 lactosamine residues, Le-FUT and Se-FUT are competing to generate an $Le^a$ or type 1H epitope ($Le^d$), respectively [46,47] (Fig. 2A). Type 1H (or Se antigen) is a substrate for the ABO gene-encoded A or B transferase, producing the A or B antigen, if one of the active alleles is present. Furthermore, type 1H can be fucosylated by Le-FUT, generating the $Le^b$ epitope. $Le^a$ and $Le^b$ structures cannot be further modified, whereas A and B antigens are again substrates to Le-FUT, resulting in $ALe^b$ and $BLe^b$ antigens [6,28,48,49]. This pathway takes place in secretory tissues and cells other than erythrocytes, since type 1 structures are highly expressed in outer epithelial layers with higher degree of differentiation, e.g. in the oral or gut mucosa, and are substrates to Se-FUT [4]. Se-FUT is active in about 80% of Caucasians, the so-called secretors (Table 1). If the Le gene is inactive, only the precursor structures $Le^c$ (type 1 precursor) and $Le^d$ (type 1H), that are classified as part of the BG collection 210 [3] , are found in plasma or on red blood cells of non-secretors and secretors, respectively [28]. Erythrocytes normally have no Se-FUT or Le-FUT expression [50], and bear ABO antigens mainly on type 2 chains, which are the preferred substrates for H-FUT [4].

On type 2 structures $Le^x$ antigens (CD15) in alpha1,3-linkage to the subterminal GlcNAc are being generated by either the Le-dependent alpha1,3/4-FUT or one of the other alpha1,3-FUTs [28] (Fig. 2B). In addition to Se-FUT, if expressed in the respective cell type, the H gene-regulated H-FUT will primarily synthesize type 2 H structures, which can be further modified by A or B transferases or alpha1,3-FUTs, incl. Le-FUT3. The action of this type of FUTs results in $Le^y$ epitopes, i.e. type 2 isomers of $Le^b$. In analogy to the above-described pathway for type 1 chains, in type 2 $ALe^y$ and $BLe^y$ will be the largest end products if all the respective glycosyltransferases are expressed in their active form [6, 28]. The terminal $Le^b$/$Le^y$, A- or B-epitopes can usually not be further modified by elongation or branching; the same applies to the subterminal $Le^a$ or $Le^x$ [6,51].

## 2.3 P1PK, FORS, GLOB and related histo-blood group antigens

The antigens of the P1PK, FORS, GLOB BGs and related collections are all GSLs (reviewed in [12, 24]). The classification of these antigens has been changed several times in the past; the current state according to the ISBT is listed in Table 1. Structurally, GSLs are composed of a lipophilic part containing a long chain fatty acid and a sphingosine anchored in the plasma membrane, and a hydrophilic glycan head group (Fig. 1). Starting from lactosylceramide either Pk antigen (globotriaosylceramide, CD77) is synthesized via the action of the A4GALT gene-encoded galactosyltransferase leading to the globo-series GSL pathway, or P1 antigen is synthesized *via* the action of the same enzyme following two other (non-BG-related) glycosyltransferases generating *neo*lacto-series GSL (Fig. 3). For P antigen production an active B3GALNT1-gene is required. The structures of the globo- and (*neo*)lacto-series GSLs act as precursors for GSL-attached ABO, Le and H/Se epitopes, as indicated in Fig. 3. If a null-allele of either A4GALT or B3GALNT1 is apparent (in very rare cases), globo-GSLs cannot be produced [52]. Furthermore, a rare variant of the A4GALT enzyme is linked to the expression of NOR antigens [53]. The FORS1 antigen was found in individuals with an activated GBGT gene, which is normally not active in humans, but in some non-primate animals [27]. Except for P1 antigen, which is RBC-specific, the expression of the other related antigens is common to many tissues and cell types, and expression levels can vary during cell cycle and differentiation [12].

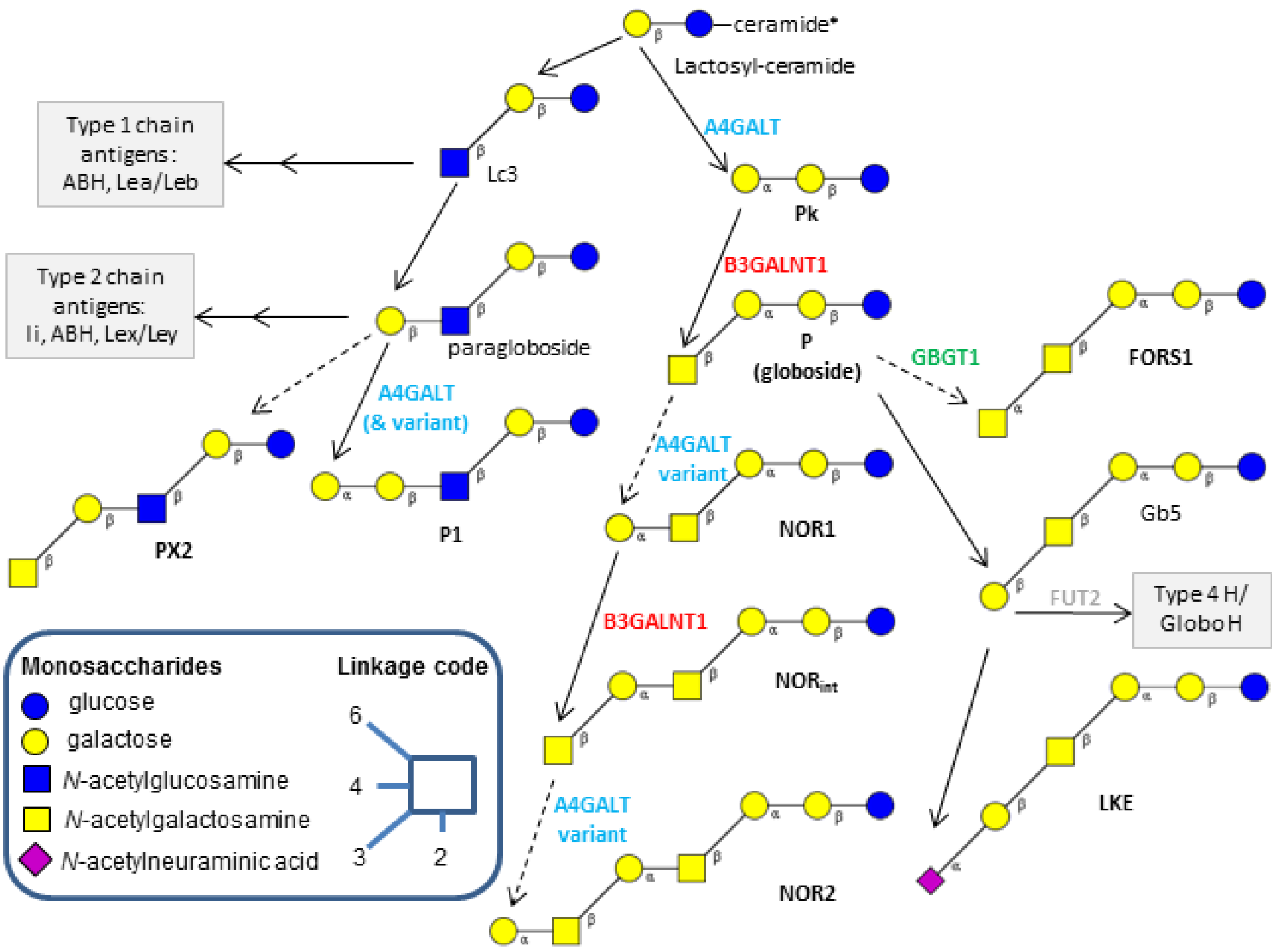

**Fig. 3 The biosynthetic pathways of the glycosphingolipid-based blood group (BG) antigens. The enzymes and resulting antigens linked to the following BGs are depicted: P1PK BG (A4GALT gene, cyan): Pk, P1, NOR; GLOB BG (B3GALNT1 gene, red): P, PX2; FORS BG (GBGT1, green): FORS1; GLOB collection 209: LKE. For full enzyme names see Table 1. *Each of the structures shown carries a beta-linked ceramide residue at the reducing end glucose, which is not depicted here for simplicity reasons. The links to the synthesis pathways of GSL-based blood group antigens of ABO, Le, and H/Se groups are also shown (grey boxes). Solid lines represent common pathways according to common glycosyltransferase gene alleles, whereas dashed lines symbolize very rare ones. Modified from [12, 24].**

## 2.4 Other glycan histo-blood group antigens

In addition to the blood group systems and collections as classified by ISBT, two more glycan histo-BGs should be mentioned here, i.e. the high incidence antigen Sd$^{a}$ and the T/Tn system. Tn antigen as part of an O-linked mucin-type glycan is primarily a substrate for T-synthase, a ubiquitously ex-pressed beta3-galactosyltransferase responsible for T antigen (O-glycan core 1) formation (for review, see [54]). T antigen is furthermore identical with type 3 chains, which can be further elongated and/or decorated by other BG antigens (Fig. 2C).

$Sd^a$ antigen is included in the high incidence 901 series according to ISBT classification due to its high prevalence in Caucasians. Similar to the discrepancy of erythrocyte vs. secretions phenotypes in Le BG, individuals having an RBC $Sd^{a-}$ phenotype may still display $Sd^a$ antigens in their secretions and especially urine (for review, see [26]). The minimal antigenic structure as shown in Fig. 2E is shared by both $Sd^a$ and CAD antigens, which are assumed to be products of the same B4GALNT gene-encoded beta4-N-acetylgalactosaminyltransferase, however, the latter resulting from a more active enzyme variant. Consequently, $Sd^{a+}$ individuals have $Sd^a$ structures primarily on N-linked glycans, whereas CAD-individuals also express these on type 3 O-linked glycans and long-chain sialyl-paraglobosides [55].

# 3 Histo-blood group phenotype vs. genotype

The genetically determined repertoire of glycosyltransferases is the basis for an individual's histo-BG phenotype. However, **genetic diversity** with the ~300 alleles found for the ABO locus and ~50 alleles for the H, Se, and Le loci each [25], together with zygosity gives rise to an enormous variation of the levels of antigen expression including weak phenotypes leading to blood grouping discrepancies [56,57]. Moreover, in pregnant women, individuals with different hematologic disorders, and especially in newborns, weak expression of histo-BG antigens on RBCs has been reported [57,58]. The age-dependency of ABO antigen expression on RBCs is connected to the expression of I antigen, which is the major precursor of ABO structures on RBCs [12]. I antigen expression in RBCs is negligible at birth and reaches the full adult level by the age of 18 months [59], giving an example of the **oncodevelopmental** nature of histo-BG antigen expression [33]. The expression levels of ABO/Se and Le antigens can also vary tremendously within an adult individual over time [60]. Regulatory mechanisms of histo-BG-related glycogene expression, such as microRNA or transcription factor expression, are now being studied [38,61].

The vast variability of the actual histo-BG phenotypes furthermore derives from the numerous **interrelations** between the respective biosynthetic pathways. This is demonstrated by the close relationships between ABO/Se and Le BGs as well as P1PK, GLOB and FORS, in addition to their precursor chains Ii and T/Tn (Fig. 2A-D). Another important modification, which is, however, not directly related to blood groups, is alpha2,3-sialylation of the terminal galactose prior to the action of a FUT on type 1 or 2 chains producing sialyl-$Le^a$ and sialyl-$Le^x$ antigens, respectively (Fig. 4). These combined histo-BG antigens are recognized for their role in the context of cancer as is discussed below. Interestingly, sialyltransferases also compete with FUTs for the same substrates, and can therefore have an impact on the expression levels of the inter-connected ABO and Le glycan BG antigens [12]. The different levels of substrate specificities have an additional impact on the overall complexity of the biosynthetic network of histo-BG glycosyltransferases: Some antigens can be synthesized by more than one en-zyme (see H/Se) and in some instances enzymes are not limited to only one type of acceptors (see chain types 1-4 for Se-FUT or the various substrates of Le-FUT), or one type of linkage (s. Le-FUT acting as alpha1,3 and alpha1,4-FUT).

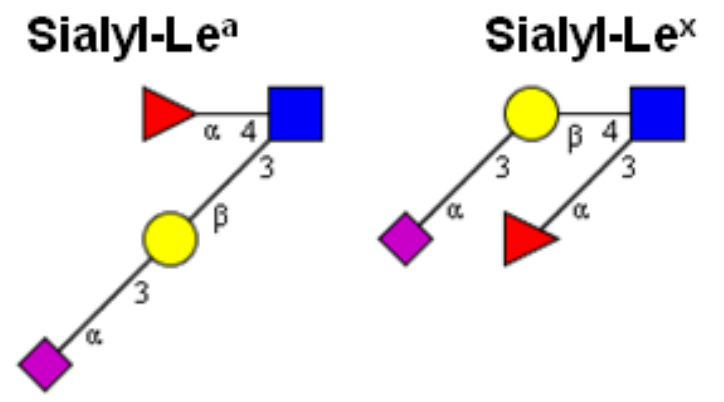

**Fig. 4 Oncodevelopmental histo-blood group antigens sialyl-Le$^a$ (CA19-9) and sialyl-Le$^x$**

Taken together, various factors can have an impact on the final phenotype, i.e. the presence of individual histo-BG glycans on a cell surface or in a body fluid. Obviously, the various associations found between BG glycans and different diseases, disease stages and health-promoting factors are making histo-BG glycans an intriguing topic in the field of personalized medicine as discussed in the following paragraphs. In Table 2 the major conclusions from selected epidemiological studies reporting on associations between different diseases and ABO, Le and Se histo-BGs are summarized.

# 4 Infectious diseases and glycan histo-blood groups

For all the common and most rare BG phenotypes on RBCs no direct causative relationship is known between the BG null-alleles and inherited diseases [5, 12]. Since glycans are known for their predominant role as recognition molecules, in particular, in infection, the association of glycan histo-BG polymorphisms with various infectious diseases is not surprising. Many vertebrate species have maintained a functional AB gene, however, in humans roughly half of the population has O genotype resulting in a non-functional A/B enzyme. ABO polymorphism has been linked to evolutionary adaptation as defense against inter- and intra-species infections, since individuals produce **antibodies against the non-self AB antigens** after the exposure to these antigens originating from e.g. pathogens [62]. **Masking of pathogen adhesion glycotopes** by other glycans is another defense mechanism suggested [63]. Whereas **microbial attachment sites** on epithelial surfaces can support colonization, histo-BG antigens on soluble glycans such as mucins or free oligosaccharides from human milk may serve as **decoy receptors** in pathogen defense [64–66] (Fig. 1).

**Table 2 Glycan histo-blood groups (BG) and selected disease associations**

| BG | Type | Disease susceptibility | Reference |
|---|---|---|---|
| **ABO** | O | *H. pylori* infection | [63] |
| | | *E. Coli* O157 infection and death | [67] |
| | | Peptic ulcer | [68] |
| | A | *S. mansoni* infection & disease severity | [69] |
| | | Gastric cancer | [68] |
| | | Overall cancer | [70] |
| | B | Salmonellosis | [71] |
| | | *E. Coli* infection | [71] |
| | Non-O | Severe malaria | [72] |
| | | Exocrine pancreatic cancer | [73] |
| | | Cardiovascular disease | [74] |
| **LE** | Inactive | Urinary tract infections | [75] |
| | | Invasive ductal breast carcinoma | [76] |
| | | Childhood asthma | [77] |
| **Se** | Inactive (non-Se) | *S. pneumonia* infection | [78] |
| | | *N. meningitidis* infection | [78] |
| | | *H. influenza* infection | [79] |
| | | Urinary tract infections | [75] |
| | | Gram-negative sepsis in premature infants | [80] |
| | | Necrotizing enterocolitis in premature infants | [80] |
| | | Gastric disease | [81] |
| | | Crohn's disease | [22,82] |
| | | Primary sclerosing cholangitis | [82] |
| | | Chronic pancreatitis | [21] |
| | | Type 1 diabetes | [83] |
| | | Breast axillary lymph nodes metastasis | [76] |
| | Active (Se) | Norovirus infection | [84–87] |
| | | Rotavirus infection | [88] |
| | | Influenza virus A & B infection | [89] |
| | | Rhinovirus infection | [89] |
| | | Respiratory syncytial virus infection | [89] |
| | | Echovirus infection | [89] |
| | | HIV infection and disease progression | [90] |

Glycan epitopes including histo-BGs have an important role in host-pathogen interactions, since glycans act as recognition sites for **bacterial** adhesins, and secondly, pathogens express surface epitopes to mimic those of the host to evade immune response, as proposed for *Helicobacter pylori* Le antigens [14]. Remarkably, *H. pylori* is able to bind to the same antigenic structures on the host's epithelial surface [91]. *H. pylori* is present in half of the world population and chronic infection is linked to gastritis, peptic ulcer and gastric cancer with a high degree of heterogeneity in disease phenotypes [92]. The bacterium is able to attach only to $Le^b$ antigen without additional A or B epitopes, and has therefore been proposed to be linked to BG O [91]. However, clinical data on the association of *H. pylori*, gastric cancer and ABO/Le BGs are contradictory (Table 2). A higher incidence of *H. pylori* in O and a lower incidence in A-individuals have been reported in multiple studies [63]. Strikingly, gastric cancer risk was higher in A BG (incidence rate ratio 1.20, 95% CI 1.02–1.42), whereas peptic ulcer risk was found to be higher in O-individuals in a large population-based study [68]. The link to Le and Se phenotype seems to be even more unclear. For instance, Se status and *H. pylori* infection have shown to be independent risk factors for gastric disease, with a higher risk in non-Se [81], although $Le^b$ (Se) seems to play a crucial role in *H. pylori* adhesion [12, 91]. On tissue level, *H. pylori* infected patients with gastric ulcer were shown to have increased $Le^a$ and loss of H and $Le^b$ expression in their inflamed gastric mucosa [7]. The authors claimed ABO and Le antigens to be good indicators for cellular alterations in the gastric epithelium.

Le- and Se-positive women were found to have lower risk for urinary tract infections as compared to Le-negative or non-Se [75]. Since adhesins from uropathogenic *Escherichia coli* recognize various globo-series GSLs including FORS1, P, and especially the sialylated LKE, antigens *in vitro* [93], a masking function of antigens competing with sialylation in secretors, i.e. $Le^b$ and AB, has been proposed [94]. B and AB individuals in a pediatric cohort were found to be more susceptible to salmonellosis and various *E. coli* types of infections (meningitis, sepsis, pyelonephritis) as compared to A or O individuals [71]. In contrast, *Shigella* infections were more prevalent in O individuals [71]. This relates to the higher susceptibility of O-individuals to infection by another Shiga-toxin related bacterial strain, i.e. *E. coli* O157 [67]. The same study found P-negative individuals to be more likely to develop severe hemolytic uremic syndrome after infection, possibly due to higher induction of TNF-alpha [67]. Pk, LKE and FORS1 histo-BG antigens have been associated with other Shiga toxin- and verotoxin-producing *E. coli* strains (STEC and EHEC) [24,95]. Moreover, Pk-knockout mice showed no reaction to verotoxin administration, whereas basal levels of Pk expression were already lethal to wild-type mice [52].

Shiga- and verotoxin-binding Pk antigen is further found in human milk, and may act as decoy receptor in breastfed infants, that are at lower risk for diarrheal diseases as compared to formula-fed ones [96]. The level of Se-positive free oligosaccharides in human milk has also been positively correlated with, e.g., protection against *E. coli*-induced diarrhea in breastfed infants [97]. Similarly, alpha-1,2-fucosylated oligosaccharides from human milk were associated with protection against *Campylobacter jejuni* diarrhea [98], possibly due to their function as decoy receptors [99]. In premature infants, non- or low-Se phenotype as well as genotype was furthermore associated with severe outcomes, such as necrotizing enterocolitis or gram-negative sepsis [80].

Expressing AB antigens on epithelia and in secretions, as is the case in Se-positive individuals, might furthermore be protective against pneumonia caused by *streptococci* [63, 78]. This pathogen secretes glycosidases acting on type 2 chain A, B and $Le^y$ structures, which are known virulence factors [100]. Secretors predominantly express type 1 chains, shielded by A and B epitopes and might therefore be protected against pathogen adhesion.

Non-secretors have also been reported being more susceptible to infections caused by *Neisseria meningitidis* and *Haemophilus influenzae* [78, 79]. Se, Le, and AB BGs seem to also play a role in *Vibrio cholerae* infection and disease severity [101,102].

Pk, P, ABH antigens and 2'- and 3-fucosyllactose, the latter two being milk oligosaccharides generated by Se- and Le-fucosyltransferases, respectively, may play a role in *Pseudomonas aeruginosa* adhesion [103,104]. Pk antigen also binds to adhesin P from *Streptococcus suis*, a pig pathogen that can cause meningitis in humans [105].

Various studies have demonstrated the stimulation of antibody titers against BG antigens under different conditions of bacterial exposure, thus, supporting the hypothesis of an immunologic defense function of BG antigen (non-)expression [12]. Another indication for the implication of ABO histo-BGs in immunological processes was given by the findings that the efficacy and the vibriocidal antibody response of cholera vaccines was lower in BG O compared to BG A individuals [106,107]. Therefore, exploring the role of anti-BG antibody response and protection following specific vaccinations is of great interest, as has been shown in a systems biology approach using glycan microarray analysis of immunoglobulin G preparations from different healthy populations [108]. The pooled antibody preparations bound to various self-antigen ligands including especially GLOB, FORS1, $Le^y$, BG A1 and BG B epitopes, but also H2 and sialyl-$Le^a$ in some cases. The authors suggested a protective role for these self-antibodies by blocking bacterial attachment sites if released on epithelial surfaces.

In the context of **viral** infections, associations of non-B as well as Se genotype and phenotype with norovirus or rotavirus infection and gastroenteritis have been reported in numerous studies (Table 2) [84-88]. In addition, a certain rotavirus genotype, P[11] that mainly infects neonates, was found to attach specifically to saliva from neonates and infants, but not adults, due to high expression of polylactosamines, i.e. the i antigen [109]. Different viral proteins from selected rotaviral genotypes were found to bind to either i epitopes or their type 1 isomers with or without internal $Le^x$ of human milk oligosaccharides in a strain- and viral protein-dependent manner [64]. Consequently, infectivity seems to be highly dependent on the virus genotype in interaction with the host ancestry/ethnicity and different histo-BGs. For norovirus, multivalent virus-like particle vaccines have therefore been proposed representing a broad set of viral genotypes and strains [110]. Interestingly, anti-histo-BG, but not immunoglobulin G or total antibody titers prior to the challenge with norovirus were positively correlated with protection against infection in placebo-recipients, while in vaccinees the frequency of severe disease was lower with higher pre-challenge anti-histo-BG antibody titers [111]. For the development of efficient vaccines against norovirus and rotavirus, individualized approaches addressing all the relevant factors such as virus genotype, the specific particles and adjuvants used in the vaccine, and their interaction with the histo-BG antigens and anti-histo-BG antibodies of the recipients will be of particular interest in the future.

O-BG was associated with resistance towards severe acute respiratory syndrome caused by coronavirus [112]. Se phenotype has been linked to other respiratory virus diseases, including influenza, rhinovirus, respiratory syncytial virus, and echovirus [89]. Non-secretor genotype and phenotype are also associated with protective aspects in HIV, such as reduced infection risk ($p = .029$) [90] and slower disease progression ($p < .001$) [113]. Likewise, Pk expression has been shown protective against HIV-1 infection *in vitro* as compared to p phenotype [114]. The role of the P1Pk BG in HIV infection is reviewed in [115]. In contrast, P antigen might be implicated in parvovirus B19 infection, shown as its primary receptor [116].

BGs have also been associated with **parasitic** diseases (Table 2), in particular, BG-O is a recognized genetic factor for a lower risk of severe malaria [72]. The high percentage of O-individuals in malaria-endemic regions supports the hypothesis for a selective advantage of having BG O [13]. Furthermore, the incidence of schistosome infection and disease severity has been linked to A-BG [69].

# 5 Non-communicable diseases and glycan histo-blood groups

Since histo-BG glycans are recognized for their role as oncodevelopmental antigens [33] and are moreover linked to tissue differentiation stage [4], it is not surprising that changes in their expression in the context of **malignancies** have been described [30,117] – and more recently also other glycosylation changes in cancer (for extensive reviews, see [15,118,119]). Moreover, epidemiological studies found a decreased risk of various types of cancers in BG O and an increased risk in BG A, as found in a meta-analysis [70] (Table 2). However, in another meta-analysis no significant association was found when comparing O vs. non-O BGs, except for exocrine pancreatic cancer (OR=0.53 and 95% CI 0.33–0.83 for BG-O vs. non-O) [73]. Further associations were found for Le and Se genotypes with breast cancer susceptibility and axillary lymph nodes metastasis, respectively [76].

In clinical samples, an increase of either incomplete/truncated precursor structures (T, Tn, Pk, H, Ii, type 1 chain) or *de novo* expressed cancer antigens, i.e. increased sialyl-Le$^{a/x}$ (Fig. 4), or FORS1, P, Le$^{b}$ (Se) and A antigens in otherwise null-allelic individuals, is observed in various cancer types depending on the cancer stage [15,120,121]. Accordingly, sialyl-Tn is a recognized tumor marker and has been proposed as a target for the design of anticancer vaccines [122], however, with limited success to date [123]. Sialyl-Le$^{a/x}$ appear to play a role in later tumor stages and especially in metastasis [15]. Notably, FUT3 (Le) and alpha2,3-sialyltransferase transcription and sialyl-Le$^{a/x}$ antigen expression was shown to be upregulated in *H. pylori* infected patients with early onset gastric carcinoma [124]. Moreover, sialyl-Le$^{a}$ (CA19-9) is elevated in various types of gastrointestinal carcinomas and is an approved prognostic marker for pancreatic cancer therapy, however, in Le-positive patients only. Since Le-negative individuals do not express an active alpha1,4-FUT, sialyl-Le$^{a}$/CA19-9 serum levels are significantly lower or even non-detectable in pancreatic cancer patients with null-allelic Le-gene [17].

Another significant glycomic BG antigen change in cancer is a loss of A and B antigens in premalignant or malignant and metastatic tissue of otherwise A- or B-positive individuals [120,125–127]. This switch has furthermore been linked to enhanced cellular motility and poor prognosis [117]. The A/B expression loss goes along with a higher abundance of the precursor structure I antigen in carcinomas from secretors, whereas cancerous tissue from non-secretors experiences a lower I antigen expression [121]. Moreover, an increased distribution of H antigen throughout all the layers of malignant tissue as compared to a more restricted pattern in normal mucosa has been found [126,128]. Analogously, alpha1,2-FUT activity from both Se- and H-gene encoded enzymes were higher in rectal carcinomas than in normal tissue [129]. Interestingly, Se/H-antigens were found in cecal tumors from Se- as well as non-Se patients, while being absent in normal tissues of non-secretors. A very recent in-vestigation of cyst fluid and tissue from ovarian tumors has shown remarkable differences in ABO BG antigen expression between different tumor stages and subtypes [130]. By using mass spectrometry average compositions of fucosylation (ABO and Le antigens) and sialylation of mucin-type glycans, the authors were able to distinguish three groups of samples: i) low-grade, low malignant potential carcinomas and benign mucinous, ii) all serous epithelial ovarian carcinomas, and iii) serous benign samples. Again, non-secretor samples were exceptions due to a genetically determined lower expression of BG epitopes on epithelia.

The described glycomic changes in cancer have a potential as therapeutic vaccine targets, including histo-BG antigens [16,131]. Moreover, ABO BG was demonstrated as a possible parameter for detecting responders to a therapeutic prostate cancer vaccine [132]. Compared to BG A/AB participants, BG O and B patients experienced enhanced survival after vaccination, especially if they developed an antibody response against the FORS1 antigen. This was explained by the authors by the presence of FORS1 in the poxviral vector used for the vaccine, showing structural similarity to the A-antigen. If verified, the conclusions drawn from this and similar studies could open an invaluable opportunity for an effective immunization strategy in a blood-group specific manner by designing vaccines with BG epitopes complementary to the BG of the recipients who would thereby develop an enhanced immune response.

In summary, malignancy-associated changes of histo-BG glycans can be different in individuals with different genetic histo-BG backgrounds and may even be tumor stage- and subtype-specific, revealing them as promising biomarkers for (early) cancer diagnostics, as vaccine targets, and as relevant factors to include in personalized oncotherapeutic approaches.

Associations between ABO and especially the non-Se phenotype or genotype with **inflammatory and autoimmune diseases** such as Crohn's disease, primary sclerosing cholangitis, chronic pancreatitis, and type 1 diabetes have been reported [21, 22, 82, 83]. In addition, childhood asthma was found to have a higher incidence in secretors with BG O and/or Le-negative phenotype [77]. A possible explanation for the higher susceptibility to Crohn's disease in non-secretors is the link via **gut microbiota**. Intestinal microbiome composition differs between healthy Se and non-Se individuals as well as between different ABO phenotypes among secretors [133,134]. Moreover, perturbations in gut microbiota metabolism and other factors of the host-microbial environment in non-secretors and FUT2 heterozygous individuals with Crohn's disease have been reported [135]. Fucosylation in the gut, which is primarily mediated by FUT2 and can be induced by certain gut bacteria *via* yet unknown mechanisms, has a central role in the host-microbe symbiosys and can suppress pathogens [136]. Interestingly, host fucose utilization by specific microbes is only down-regulated in non-Se mice if fed a glucose-rich, polysaccharide-deficient diet, demonstrating an intense interaction between host genetics, diet and gut microbiome [137]. Notably, gastrointestinal malignancies, in particular colorectal cancer, have also been linked to gut microbiota composition [138].

Intriguingly, a previously unrecognized regulatory role of histo-BG antigens in inflammatory adhesion processes has been proposed due to a strong correlation of the ABO locus with plasma concentrations of the soluble Intercellular Adhesion Molecule 1 (sICAM-1) [139]. Moreover, the ABO locus was associated with diabetes type 2 and E-selectin levels, besides sICAM-1 levels [140]. In addition, a lower risk of **cardiovascular and associated diseases** has been found for BG O individuals and linked to BG antigens on platelet glycoconjugates, as recently reviewed [74]. Interestingly, BG O individuals have lower levels of the blood clotting factors von Willebrand and FVIII due to their shorter plasma half-life [141].

In addition to cancer-related changes in Ii antigen expression mentioned above, a rare disease association has been reported for the adult i phenotype. A correlation was found with congenital cataracts in certain kindreds [44]. This has been putatively explained by a direct effect of the lack of branched type 2 polylactosamines resulting from a certain null-allelic gene variant of the I-antigen producing enzyme in the human lens epithelial cells [23, 38, 44].

Some indications do exist that variants of the T-synthase gene generating T antigen are associated with susceptibility to IgA nephropathy [142].

A downregulation of the B4GALNT2 gene resulting in strong reduction of $Sd^a$ antigen expression in the colon and a concomitant increase of sialyl-$Le^x$ expression is observed in colon cancer and might offer a promising opportunity in cancer therapy in the future [143].

## 6 Conclusions and future perspectives

As defined by the US Food and Drugs Administration, personalized medicine is the "*tailoring of medical treatment to the individual characteristics, needs, and preferences of a patient during all stages of care, including prevention, diagnosis, treatment, and follow-up*" [144]. As briefly demonstrated by the example of the interaction of an individual's Se status and gut microbiota, histo-BGs might play a significant role already at the early stage of disease prevention. Another example is the various histo-BG glycans in human milk, which are suggested to provide double protection to the breastfed child *via* their prebiotic and their antimicrobial activity in the gut [145]. The decoy function of milk glycans, in particular the highly abundant free oligosaccharides and mucin glycans, offers a yet underexamined possibility for a novel preventive and also therapeutic antibiotic strategy. Emerging high-throughput analytical technologies such as glycan microarrays, multiplexed capillary electrophoresis, or matrix-assisted laser desorption/ionization (MALDI) mass spectrometry (MS) will pave the way for the discovery of such novel bioactive compounds [64,65,146,147].

Traditionally, BG phenotyping is performed via targeted immunological assays using antisera, lectins, or monoclonal antibodies with or without staining on body fluids or tissues to determine the specific antigenicities [41,121,130,148,149]. An immunoassay is also used for the determination of CA19-9 (sialyl-$Le^a$) levels in clinics [17]. Genotyping, glycosyltransferase transcript analysis as well as standard proteomic techniques are currently being increasingly used also in the context of BG-typing [27, 130], and do therefore comply with the requirements of modern personalized medicine. Furthermore, they might be useful for patient stratification in settings where the genetic histo-BG background in combination with the natural anti-BG antigen antibodies plays a role, such as vaccine efficacy studies. The reverse ABO-typing method developed by Muthana and colleagues by using only 4 µL of serum with glycan arrays was proven as particularly useful since it enabled retrospective BG typing without having whole blood samples available [132].

Our current knowledge on the significance of histo-BGs in the context of disease as summarized in sections 4 and 5 is thus far mainly based on observational studies stratifying patients and controls by their blood type and showing associations with disease or disease stages (Table 2). Information on the underlying mechanisms is very limited to date, possibly due to the extremely reductionistic approach that emerged from transfusion medicine, i.e. grouping individuals into a few categories deduced from their serological phenotype (Table 1 and 2). However, the knowledge of a patient's RBC pheno-/genotype or glycosyltransferase expression is only indicative for the final, complex glycomic profile [62], especially if taking into account the vast tissue- and disease-specific variations of the histo-BG antigens, their precursors and conjugates, as well as interactions between different BGs, as discussed in sections 2 and 3. Thus, simultaneous mapping of glycans from different histo-BGs, next to the information on their carriers would provide the increased degree of comprehensiveness that is required to investigate the mechanisms behind the BG-to-disease associations presented above.

Modern glyco-analytical techniques, such as MS and liquid chromatography, are in principle capable to provide this type of information on multiple levels:

1) (semi-)quantitative determination of terminal glycosylation including histo-BG glycans as well as sialylation, the levels of which appear to be inter-related;
2) mapping of the overall glycome by furthermore providing information on the core glycan structures bearing the terminal BG glycans, i.e. N- or O-glycans, or the glycan portion of the GSL-based BG structures;
3) determination of the protein/peptide or lipid carrier.

Thus, instead of just assigning samples to a limited set of categories as based on traditional or genetic BG typing, advanced glyco-analytics will enable an unprecedented insight into the vast complexity and variety of structures (Fig. 1) reflecting the underlying complex biological processes. Therefore, glycomic, together with glycoproteomic and glycolipidomic technologies should be further refined to allow unambiguous (isomer-specific) structural characterization and optimally quantitation of histo-BG glycans that have high potential as diagnostic or prognostic biomarkers in the context of both infectious and communicable diseases. Currently applied glyco-analytical techniques in the field of biomarker discovery from cancer tissues are presented in [150]. Major progress in glycomic approaches distinguishing linkages and/or different monosaccharide species has been made recently in the field of N- and O-glycomics [151,152]. However, glycomics of histo-BGs has been neglected in the last years except for some few, rather targeted approaches [153–155]. In a very recent study traditional, targeted techniques in combination with liquid-chromatography-MS/MS were successfully applied showing great potential of mucin-type histo-BG glycans for ovarian tumor staging and classification not only from tissues, but also from cystic fluid [130]. Glycomic approaches on only minute amounts of preferably non-invasively acquired patient samples, such as saliva, tears, milk, urine, or feces [66,156–159], will be of particular use in the context of early diagnosis and population-wide screening.

The discovery of novel functions of histo-BGs by applying high-end technologies is warranted. Interestingly, the spatial organization of ABO antigen patches on RBCs from A, B, and O individuals was shown to differ in BG-specific manner resulting in higher or lower binding of sialic acid-binding proteins (Siglecs) to sialic acid-rich clusters that are co-expressed nearby the ABH regions [160]. The authors claimed having found a novel function of ABH antigens, i.e. as indirect stabilizers of other glycan-protein interactions. Further research is warranted to reveal the function relevance of such multi-compound complexes involving the concerted binding of glycans in a multivalent manner.

Finally, a combination of the different OMICs techniques including glycomics will likely enhance the performance of the diagnostic and prognostic approaches [15] and pave the way for a successful translation into clinics. As a first example, in a pioneering combined genomics-glycomics study potential epigenetic regulatory factors of Le-FUT (FUT3), next to other FUTs, were proposed on the basis of plasma protein N-glycome [161]. Studies combining genomic, proteomic, metabolomic and glycomic data containing histo-BG antigen information of different body fluids and/or tissues could help revealing the mechanisms behind the above described disease associations and may provide improved disease biomarkers as well as therapeutic targets in the future.

## Conflict of interests statement

The authors declare no conflict of interest.

## Acknowledgements

This work was supported by the European Union with the Seventh Framework Programme HighGlycan project, grant number 278535, and the Horizon2020 GlyCoCan project, grant number 676421.